\documentclass[prb,twocolumn]{revtex4}% Physical Review B

\usepackage{graphicx}% Include figure files
\usepackage{dcolumn}% Align table columns on decimal point
\usepackage{bm}% bold math

\begin{document}

\preprint{APS preprint}

\title{Transport in two-dimensional electron gas narrow channel with a magnetic field gradient}% Force line breaks with \\

\author{Masahiro Hara}
\email{hara@issp.u-tokyo.ac.jp}
\author{Akira Endo}
\author{Shingo Katsumoto}
\author{Yasuhiro Iye}

\affiliation{%
Institute for Solid State Physics, University of Tokyo, Kashiwa-shi, Chiba, 277-8581, Japan
}%

\date{\today}% It is always \today, today,

\begin{abstract}
We have investigated distinctive transport phenomena in two-dimensional electron gas (2DEG) narrow channel subjected to a large magnetic field gradient, a unique situation created in a ferromagnet/2DEG hybrid structure. The so-called snake orbits which propagate in the direction determined by the sign of field gradient, give rise to conductivity enhancement and rectification effect with respect to the direction of DC bias current.
\end{abstract}

\pacs{73.23.Ad,73.50.Jt,75.60.-d}% PACS, the Physics and Astronomy
\maketitle

Mesoscopic ferromagnet/two-dimensional electron gas (2DEG) hybrid structures attract much interest not only as an experimental stage of novel magnetotransport phenomena but also as a prototype of future magneto-electronics devices. A tailored magnetic field landscape can be applied to the 2DEG by an appropriately microfabricated ferromagnetic structure. Ballistic electron motion in the high mobility 2DEG is strongly modified by such local magnetic field\cite{izawa,ye,nogaret2,kato,kubrak,vancura,hara}. Recently, Nogaret {\it et al.}\cite{nogaret} reported a resonance peak in magnetoresistance of a device with a ferromagnetic strip deposited above the center line of the channel due to the so-called snake state which corresponds to electron trajectories trapped at the boundary between positive and negative magnetic field. In this work, we investigate different aspects of electron transport in a similar hybrid structure, in an experimental setup which allows us to precisely control the magnetization of the mesoscopic ferromagnet by a cross-coil magnet system.

Our samples were fabricated from a GaAs/AlGaAs single-heterojunction wafer grown by molecular beam epitaxy. The density and mobility of the 2DEG before processing were ${\rm 3.1\times 10^{15}\ m^{-2}}$ and ${\rm 67\ \ m^{2}/Vs}$, respectively. The electron mean free path was ${\rm 6.1\ \mu m}$. The depth of the 2DEG plane from the sample surface was 65\ nm. A narrow channel of 2DEG was formed by chemical wet etching with geometrical width 1.5 $\mu$m (sample $\#$1) and 1.8 $\mu$m (sample $\#$2). The physical width of the 2DEG channel is somewhat smaller due to depletion at both edges. The separation between two voltage probes to measure the longitudinal resistance $R_{xx}$ of the 2DEG narrow channel was 12\ $\mu$m. A cobalt strip of thickness 75\ nm and width 0.5\ $\mu$m was placed on top along the center line of the channel. The transport measurements were carried out using a low-frequency AC lock-in technique at an excitation current of 50\ nA for the 2DEG and 300\ nA for the cobalt wire. 

\begin{figure}[htbp]
\begin{center}
\includegraphics[width=\linewidth]{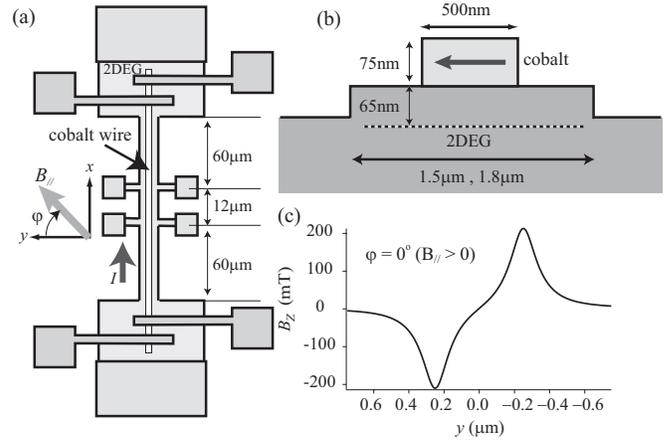}
\caption{(a) Schematic diagram of the sample configuration. Applying an in-plane magnetic field in $+y$ direction, electron trapped in snake state propagates in $+x$ direction. (b) Cross-section of the narrow channel. The width of the 2DEG channel is actually smaller than the geometrical width because of the depletion of 2DEG near the sample edge. (c) Calculated magnetic field profile $B_z$ at the 2DEG plane with the in-plane magnetic field applied perpendicular to the channel direction ($\varphi = 0^{\circ}$). The calculation was made by assuming the saturation magnetization of cobalt to be 1.8\ T.}
\label{sample}
\end{center}
\end{figure}

The device was fabricated in such a way that the electronic transport in the cobalt wire itself could be measured simultaneously as that in the 2DEG wire, in order to monitor the magnetization process of the former. Measurements were done at temperatures between 1.3\ K and 14.2\ K. Since the temperature was far below the Curie temperatures of cobalt, the magnetic behavior of the cobalt strip was temperature independent. Use of a cross-coil magnet system consisting of a 7\ T split coil and a 1\ T solenoid enabled us to control two components of magnetic field independently. By applying an in-plane magnetic field to magnetize the cobalt wire in the direction perpendicular to channel direction ($\varphi=0^{\circ})$, a spatially varying magnetic field is generated inside the 2DEG channel. Here, $\varphi$ is the azimuthal angle of the in-plane magnetic field defined in Fig.\ref{sample}(a). Fig.\ref{sample} (c) shows a magnetic field profile across the 2DEG channel calculated by a simple magnetostatic model. The maximum amplitude of the normal component $B_z$ of the stray field for this structure is about 0.2\ T, and can be controlled by azimuthal angle $\varphi$ of the applied in-plane magnetic field. 

\begin{figure}[htbp]
\begin{center}
\includegraphics[width=\linewidth]{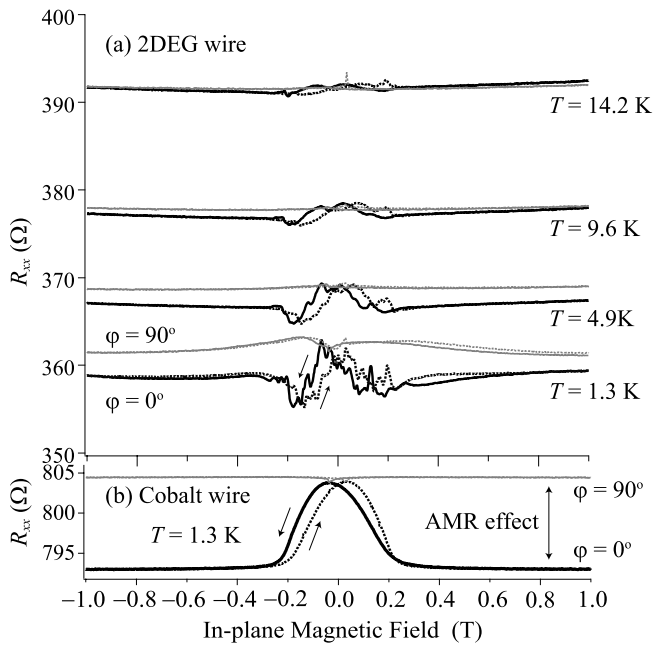}
\caption{(a) Resistance behavior of 2DEG narrow channel for sample $\#$2 sweeping in-plane magnetic field in two different directions at $T$ = 1.3 K, 4.9 K, 9.6 K, and 14.2 K. (b) Simultaneously measured resistance of the cobalt wire at $T$ = 1.3 K.}
\label{ips}
\end{center}
\end{figure}

\begin{figure}[htbp]
\begin{center}
\includegraphics[width=\linewidth]{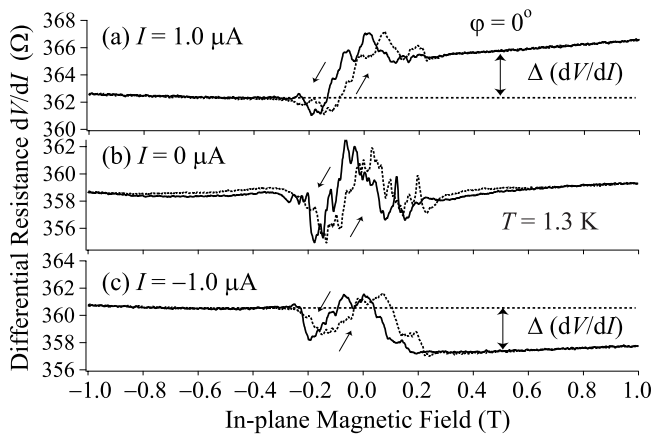}
\caption{Differential resistance of sample $\#$2 as a function of in-plane magnetic field for $\varphi = 0^\circ$ under DC bias current $I$ = $1\ \mu$A, $0\ \mu$A, $-1\ \mu$A at $T$=1.3 K. The change in the resistance between positive and negative in-plane field defined as $\Delta(dV/dI)$ alters the sign depending on flow direction of the DC bias current.}
\label{biasmr}
\end{center}
\end{figure}

We first describe the control and monitor of the magnetization of the cobalt strip by the external magnetic field. The sample was precisely aligned in the cross-coil system by monitoring the Hall signal so that the two components of magnetic field were both parallel to the 2DEG plane. The in-plane field $B_{\parallel}$ was thereafter used solely to control the magnetization of the cobalt strip without affecting the transport in 2DEG by itself. (The parallel field magnetoresistance of the 2DEG was negligible in the field range of the present work, as verified by samples without ferromagnetic structure.) Two curves of magnetoresistance of the cobalt wire measured with the in-plane magnetic field applied perpendicular $(\varphi = 0^{\circ})$ and parallel $(\varphi = 90^{\circ})$ to the current direction are shown in Fig.\ref{ips}(b). They show a typical anisotropic magnetoresistance (AMR) effect reflecting the magnetization process of the cobalt strip with some hysteresis. This particular set of data was taken at 1.3\ K, but the behavior is identical throughout the present temperature range. Due to the shape anisotropy, the easy direction of magnetization was along the length of the strip. For $\varphi = 90^{\circ}$, magnetization reversal occurs in the field range $\sim \pm $0.03\ T. When an in-plane magnetic field higher than this value was applied to the cobalt strip in this direction $(\varphi = 90^{\circ})$, the stray field was turned off.

In the magnetic field range where magnetization reversal occurs, the amplitude of the stray field pattern changes, and the corresponding change in the resistance of the 2DEG wire takes place. Figure \ref{ips}(a) shows the resistance of the 2DEG wire as a function of the in-plane magnetic field for $\varphi = 0^{\circ}$ and $\varphi = 90^{\circ}$ at four different temperatures. The overall vertical shift reflects the temperature dependence of resistivity governed by the acoustic phonon scattering.

For $\varphi = 0^{\circ}$, reproducible aperiodic fluctuation in resistance is observed in the field range where the magnetization and hence the stray field changes. The fluctuation amplitude diminishes at higher temperatures. This fluctuation cannot be attributed to discontinuous jumps of magnetization in its reversal process, because the magnetization process is smooth and temperature independent as witnessed by the magnetoresistance of the cobalt wire. The resistance fluctuation phenomenon is attributed to universal conductance fluctuation (UCF) effect of the narrow 2DEG wire. This interpretation is supported by the following observations. The amplitude of conductance fluctuation is $\sim 0.1e^2/h$  at 1.3\ K which falls in a reasonable range as UCF. Similar resistance fluctuation is observed as a function of uniform perpendicular magnetic field. The phase relaxation length at the lowest temperature is estimated as $\sim 2\ \mu$m from the amplitude analysis\cite{beenakker}, which is comparable to the sample size. By use of the magnetization curve, the horizontal axis of Fig.\ref{ips}(a) can be translated into the average (absolute) value of stray magnetic field. The values of correlation field for the aperiodic fluctuations are then found to be comparable between the stray field case and the uniform perpendicular field case.

 Comparison of the magnetoresistance curves for $\varphi = 0^{\circ}$ and $\varphi = 90^{\circ}$ in Fig.\ref{ips}(a) tells that the conductance of the 2DEG narrow channel becomes higher in the presence of magnetic field gradient. This is attributed to the contribution of snake orbits in the former case. In ballistic or quasi-ballistic transport regime such as the present case, diffuse scattering at boundaries plays an essential role. When snake states are created by gradient magnetic field, electrons trapped in those states do not suffer from boundary scattering leading to conductivity enhancement. 
 
 It is instructive to compare this behavior with the case of uniform magnetic field. In a narrow wire under a low uniform magnetic field, all electron trajectories are directed towards the boundaries causing a positive magnetoresistance. At higher magnetic field where the cyclotron diameter becomes less than the wire width, boundary scattering is suppressed resulting in a negative magnetoresistance.
 
 The extra contribution of snake states to conductivity diminishes with increasing temperature as seen in Fig.\ref{ips}. At higher temperatures, frequent electron-phonon and electron-electron scattering events tend to prevent electrons from staying in the snake orbits. 

The role of snake states is more conspicuous in the non-ohmic regime. Energy dispersion in the narrow channel under gradient magnetic field is asymmetric with respect to the current direction, {\it i.e.}, the time-reversal symmetry is broken\cite{muller,badalyan}. In a classical picture, electrons in the snake state propagate along the direction determined by the sign of the field gradient. In order to study the effect of the asymmetry, we measured a differential resistance under DC bias current with modulated AC component 50 nA. Fig.\ref{biasmr} shows the differential resistance as a function of in-plane magnetic field in the direction perpendicular to the current. Under finite DC bias current, we observed the change in the resistance $\Delta (dV/dI)$ as defined in Fig.\ref{biasmr} between positive and negative in-plane field for $\varphi = 0^{\circ}$. The sign of $\Delta (dV/dI)$ was inverted by changing the direction of DC bias current. The resistance is smaller when the current carrying direction of the snake states is the same as the DC bias current. For $\varphi = 90^{\circ}$, such effect is absent.

\begin{figure}[htbp]
\begin{center}
\includegraphics[width=\linewidth]{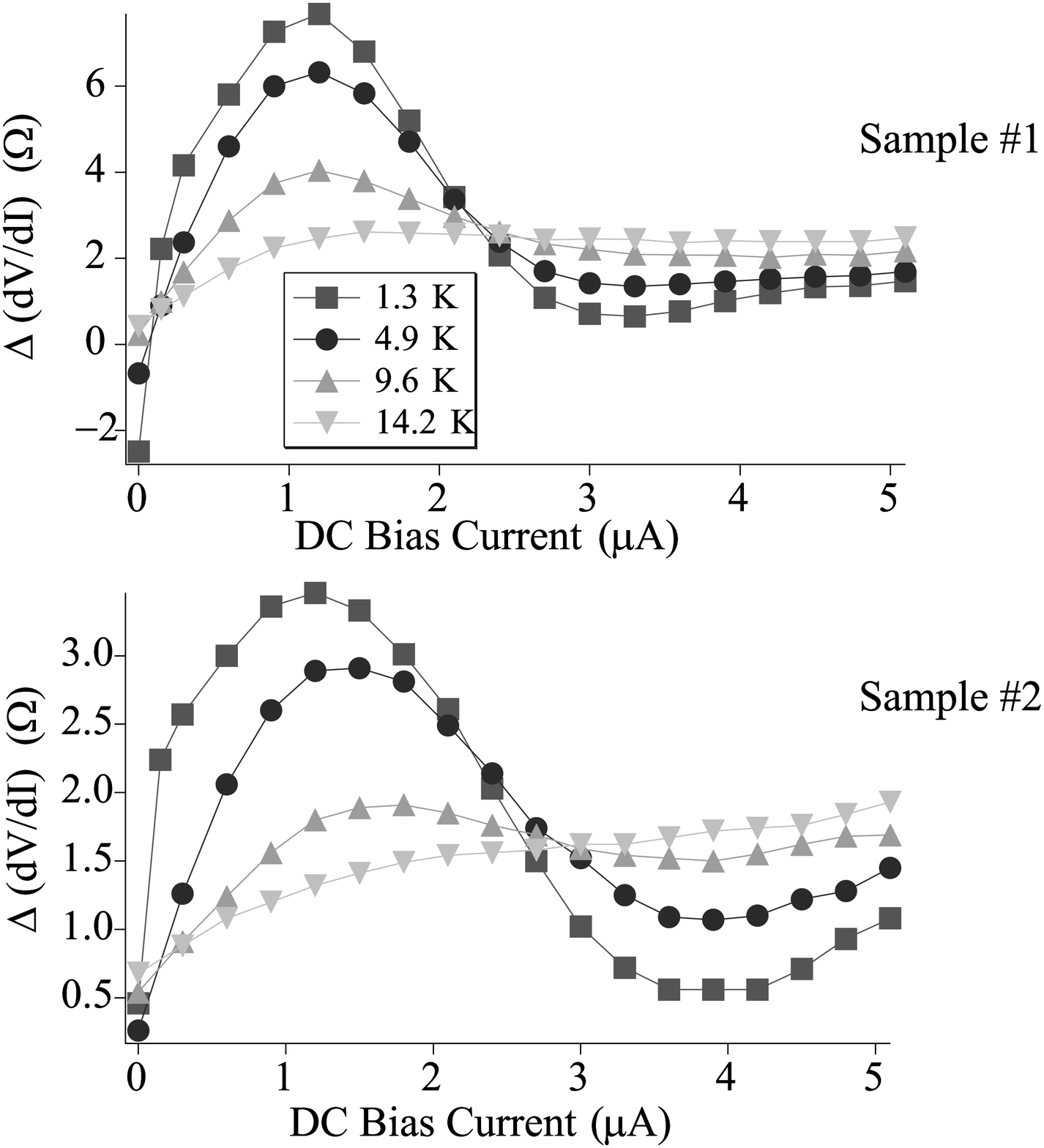}
\caption{Change in the differential resistance between positive and negative in-plane magnetic field as a function of DC bias current at $T$ = 1.3 K, 4.9 K, 9.6 K, and 14.2 K. The resistance under zero bias of sample $\#$1 and $\#$2 at $T$ = 1.3 K are about 610\ $\Omega$ and 360\ $\Omega$, respectively.}
\label{bdtd}
\end{center}
\end{figure}

\begin{figure}[htbp]
\begin{center}
\includegraphics[width=\linewidth]{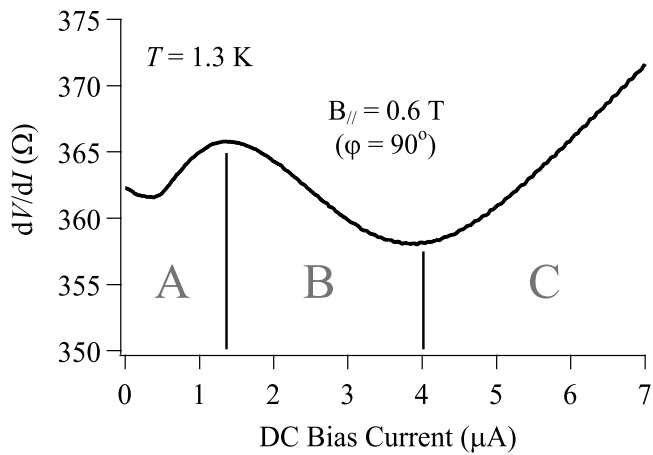}
\caption{Differential resistance as a function of DC bias current for sample $\#$2 at $T$ = 1.3 K. The gradient magnetic field was turned off by applying the in-plane magnetic field in the current direction $\varphi =90^{\circ}$. }
\label{diff}
\end{center}
\end{figure}

 A similar rectification effect has been reported by Lawton {\it et al.}\cite{lawton}. They have proposed that the effect is due to anisotropic electron-phonon interaction arising from asymmetry in the energy dispersion of 2DEG subjected to a gradient magnetic field, which becomes conspicuous when the electron temperature is raised by bias current\cite{kibis}. Here, we propose an alternative mechanism for the rectification effect in the non-ohmic regime which we believe is important in the present device structure. 
 
 In order to gain insight into the mechanism of the observed effect, we measured the bias current dependence of $\Delta (dV/dI)$ for different settings of bath temperature. Figure \ref{bdtd} shows the result for the two samples. The value of  $\Delta (dV/dI)$ increases with increasing DC bias up to $3\ \mu$A, and then decreases at higher current bias. The rectification effect diminishes with increasing temperature. We estimated the electron temperature $T_{\rm e}$ at finite DC bias currents from damping of the Shubnikov-de Haas amplitude measured at 1.3 K under uniform perpendicular magnetic field\cite{ando}. At a DC bias of $2\ \mu$A, for example, the temperature difference between the 2DEG and lattice system was estimated to be about 7 K for both samples. At an elevated electron temperature $T_{\rm e}$, the electron-electron scattering rate $1/\tau_{\mathrm{e}\textrm{-}\mathrm{e}}$ in 2DEG increases according to $1/\tau_{\mathrm{e}\textrm{-}\mathrm{e}}\propto T^2\log T$ (Ref.\onlinecite{giuliani}). The e-e scattering mean free path defined as $\ell_{\mathrm{e}\textrm{-}\mathrm{e}}=v_{\rm F}\tau_{\mathrm{e}\textrm{-}\mathrm{e}}$ is calculated to be $3.2\ \mu$m for $T_{\rm e}$ = 10 K. Such value is smaller than the the bulk mean free path 6.1\ $\mu$m due to impurity scattering.
 
 Increased electron temperature also modifies the electron-phonon (e-ph) scattering rate $1/\tau_{\mathrm{e}\textrm{-}\mathrm{ph}}$ by increase of phonon emission rate with little change in phonon absorption rate governed by $T_{\rm L}$. Thus the e-ph scattering rate replacing bath temperature $T$ with electron temperature $T_{\rm e}$ may be used as an upper limit of $1/\tau_{\mathrm{e}\textrm{-}\mathrm{ph}}$ for hot electron system. From the temperature dependent resistivity of the 2DEG, the e-ph mean free path at $T_{\rm e}$=10\ K is estimated to be longer than 80\ $\mu$m. Thus, in the present experimental situation, e-e scattering is more important as momentum randomizing scattering than e-ph scattering.

 It is well known that electron-electron scattering does not contribute to resistance in bulk 2DEG since the total momentum of electrons involved in scattering is conserved. However, in the case of 2DEG narrow channel where diffuse scattering at the side edges is important, a combined effect of e-e scattering and diffuse boundary scattering can give rise to enhanced resistance at elevated electron temperatures. Such an effect is studied by Molenkamp and de Jong\cite{molenkamp}. Figure \ref{diff} shows the differential resistance $dV/dI$ as a function of DC bias current in the absence of magnetic modulation for sample $\#$2. The behavior is in good agreement with those reported in Ref.\onlinecite{molenkamp}. As the DC bias current is increased from zero, the resistance initially increases. In this region denoted as (A), the e-e scattering length $\ell_{\mathrm{e}\textrm{-}\mathrm{e}}$ is still much larger than the channel width. Electrons suffer relatively infrequent e-e scattering events, but they are effective in enhancing the probability of boundary scattering, so that the resistance is increased. At a higher bias current (region B), the e-e scattering length becomes comparable or smaller than the channel width, frequent e-e scattering events suppress the probability of electrons hitting the boundaries, which results in resistance decrease. Further increase of bias current increases the lattice temperature (region C), causing increases of resistance.
 
 With the above situation in mind, we model the rectification effect in the presence of magnetic field gradient as follows. In the presence of magnetic field gradient, the current flow parallel to the direction of snake orbits is collimated along the center line of the channel. On the other hand, current flow antiparallel to the direction of snake orbits is predominantly carried by electrons near the edges. In other words, electrons moving in the $+x$ and those in the $-x$ directions are separated from each other in the $y$ direction, the former near the center and the latter near the edges. With increasing current bias and hence $T_{\rm e}$, the e-e scattering rate increases. However the effectiveness of the e-e scattering as enhancer of boundary scattering significantly differ between the oppositely moving electrons. The electrons trapped in the snake states along the center line of the channel are expected to be more immune to the combined effect of e-e scattering and boundary scattering. This gives a mechanism for the rectification effect at finite bias current.
 
 Our model is supported by an observation from comparison of the two samples with different width that the maximum value of $\Delta (dV/dI)$ is correlated with the channel resistance $R$ while the increase of $T_{\rm e}$ with DC bias current is nearly the same. For the narrower channel device (sample $\# 1$) $\Delta (dV/dI)=7.68\ \Omega$ and $R=610\ \Omega$, while they are $3.46\ \Omega$ and $360\ \Omega$ for the wider channel device (sample $\# 2$). It is consistent with the picture that diffuse boundary scattering is responsible for the effect. When the bias current is further increased to make the e-e scattering length less than the channel width, frequent randomization of electron momentum overwhelms the ballistic electron dynamics, resulting in disappearance of the rectification effect.
 
 In conclusion, we have observed asymmetric transport and rectification effect in a narrow 2DEG wire associated with the so-called snake state under a spatial gradient of magnetic field. We propose the combined effect of e-e scattering and diffuse boundary scattering as a major source of the asymmetric transport\cite{model}.

This work was supported by a Grant-in-Aid for Scientific Research ($\#$13304025) and a Grant-in-Aid for COE Research ($\#$12CE2004) from the Ministry of Education, Culture, Sports, Science and Technology (MEXT) Japan. One of the authors (M.H) acknowledges support by the Japan Society for the Promotion of Science (JSPS) fellowship.

\end{document}